\UseRawInputEncoding
\documentclass[twocolumn,openany,openright,amsmath,amssymb,superscriptaddress]{revtex4-2}
\usepackage[english]{babel}

\usepackage{float}
\usepackage{latexsym} 
\usepackage{amsmath,amsthm}
\usepackage{ifpdf}
\usepackage{epstopdf}
\usepackage{subfig}
\usepackage{soul}
\usepackage{dcolumn}
\usepackage{bm}
\usepackage{braket}
\usepackage{wrapfig}
\usepackage[dvipsnames]{xcolor}
\usepackage{color}
\usepackage{hyperref}

\usepackage{graphicx}

\newcommand{\beq}{\begin{equation}}
\newcommand{\eeq}{\end{equation}}
\newcommand{\barr}{\begin{eqnarray}}
\newcommand{\earr}{\end{eqnarray}}
\newcommand{\bseq}{\begin{subequations}}
\newcommand{\eseq}{\end{subequations}}

\newcommand{\expectation}[3]{\langle #1|#2|#3\rangle}

\newcommand{\vett}[1]{\textbf{#1}}
\newcommand{\uvett}[1]{\hat{\textbf{#1}}}

\usepackage[normalem]{ulem}

\begin{document}

\title{Nonlinear Quantum Optics in Epsilon-Near-Zero Nanospheres: a Macroscopic Quantum Electrodynamics Approach}
\author{Sonia Alipour}
\affiliation{Faculty of Engineering and Natural Sciences, Tampere University, Tampere, Finland}

\author{Marco Ornigotti}\email{Corresponding author: marco.ornigotti@tuni.fi}
\affiliation{Faculty of Engineering and Natural Sciences, Tampere University, Tampere, Finland}
\begin{abstract}
\noindent We present a theoretical framework for handling quantum nonlinear processes in epsilon-near-zero (ENZ) nanospheres, based on macroscopic quantum electrodynamics and the resonant states (RSs) expansion of the Green's tensor in nanocavities. Using third-harmonic generation (THG) as a case study, we derive analytical expressions for the generation efficiency, and the quantum fluctuations of the generated nonlinear field, as a function of different quantum states of the impinging pump field. Finally, we discuss how to generalise this theoretical framework to arbitrary quantum nonlinear processes in photonic nanostructures.
\end{abstract}

\maketitle
\section{INTRODUCTION}
Quantum properties of light have attracted considerable attention in the past decades, due to their fundamental importance in quantum optics and their relevance for emerging photonic technologies  \cite{khanbekyan2003input,o2009photonic,loudon2000quantum}. Advances in single-photon sources  \cite{PhysRevA.69.032305,Cui:05}, entangled-photon generation \cite{doi:10.1126/sciadv.abq4240,GIRVIN2004591}, quantum communication \cite{krenn2016quantum}, quantum computation \cite{romero2024photonic,Cui:05} and integrated quantum photonic platforms heavily rely on a rigorous description of the interaction between light and matter at the quantum level  \cite{wang2020integrated}. The presence of boundaries between media, for example, gives rise to fundamental changes to the structure of the electromagnetic vacuum, resulting in pivotal applications such as 
%
%
spontaneous-emission engineering \cite{baldo2025dynamics,WIESE200221}, the Purcell effect  \cite{VENKATAPATHI20121705,Wang_2021,abrantes2025spontaneous}, Casimir interactions  \cite{Milonni2009,shelden2026casimir}, and cavity quantum electrodynamics \cite{Northup_2012,Walther_2006}. Accounting for these effects in dispersive and lossy environments, such as epsilon-near-zero (ENZ) nanocavities, requires a rigorous quantum theory of the electromagnetic field capable of handling the presence of dispersion and losses in a consistent, causal manner  \cite{Scheel_2006,knoll2000qed,bechler1999quantum}. 

Several theoretical approaches have been developed in the past to describe QED of linearly scattering media  \cite{PhysRevA.46.4306,knoll2000qed,Scheel_2006,PhysRevA.57.4818} including the Hopfield model  \cite{PhysRev.112.1555}, the Huttner-Barnett model  \cite{PhysRevA.46.4306}, and the Bechler model  \cite{bechler1999quantum}, based on path integrals and extendable to inhomogeneous media \cite{PhysRevA.70.013816}. From a macroscopic perspective, these models all converge into the Green's tensor quantization framework introduced by Gruner and Welsch in the 90s  \cite{PhysRevA.53.1818} ad then adapted to various geometries  \cite{knoll2000qed,PhysRevA.53.1818,PhysRevA.54.1661} and nonlinear optics  \cite{Scheel_2006,StefanScheel_StefanBuhmann_2008,PhysRevLett.96.073601}.
%
%
Despite these efforts, a consistent description of resonant nonlinear optical interactions in realistic dispersive and absorbing nanophotonic systems, especially for open optical resonators, remains challenging, and only recently the Green's tensor quantization framework was used to investigate single photon dynamics in nonlinear, ENZ nanocavities  \cite{9yyx-s3m9,PhysRevResearch.5.043228}.

One of the main advantages of this framework is the fact, that macroscopic QED preserves the canonical commutation relations between electric and magnetic fields, while at the same time encoding losses in the form of vacuum fluctuations, which, thanks to the fluctuation dissipation theorem, are proportional to the imaginary part of the Green's tensor of the system under investigation  \cite{StefanScheel_StefanBuhmann_2008} or, equivalently, its local density of states  \cite{Novotny_Hecht_2012}. Nowadays, several methods, such as quasi-normal mode (QNM) solvers  \cite{Kristensen:20} or resonant state (RS) analysis  \cite{Muljarov_2010,PhysRevA.90.013834,PhysRevB.101.045304}, are available for calculating the Green's tensor of a given photonic nanostructure of arbitrary geometry and size. Recently, moreover, it has been shown how a quantisation scheme based on QNMs can be employed to describe the properties of both lossy structures, like ENZ nanocavities, as well as dielectric nanocavities, for which the losses are prevalently radiative \cite{PhysRevLett.122.213901}. Pairing these methods with the rigorous manner in which macroscopic QED handles nonlinearities at the quantum level  can therefore create a novel platform for studying quantum optical effects in ENZ materials and, in particular, nanostructures. 
%

ENZ materials, in fact,  have attracted significant attention because of their exceptional electromagnetic response near the ENZ frequency  \cite{reshef2019nonlinear,Wu:21,Koivurova_2020,liberal2017near, https://doi.org/10.1002/adom.201701292}, strong electric-field enhancement \cite{PhysRevB.87.035120, Li:20, https://doi.org/10.1515/nanoph-2020-0490,Reddy:20}, slow-light effects \cite{PhysRevA.87.053853}, and enhanced nonlinear optical responses  \cite{reshef2019nonlinear,10.1063/5.0240990}. For these reasons, the potential of using ENZ materials for a broad range of nanophotonic applications has been investigated extensively in the past years, including optical switching \cite{9144401,https://doi.org/10.1002/adma.201700754}, modulation \cite{Wood:18,Vasudev:13}, frequency conversion \cite{Khurgin:20,Zhou:19}, nonlinear high harmonic generation \cite{yang2019high,Tian:21}, and on-chip light sources \cite{vertchenko2019epsilon}. Third-harmonic generation (THG), for example, can be drastically enhanced by the presence of the local field enhancement near the ENZ region of the material \cite{article, 10.1063/5.0240990}

Motivated by all this, in this work we present a quantum theoretical framework 
which combines the 
%
%
the macroscopic QED approach of Gruner and Welsch with the RS expansion of the Green's tensor. In particular, we derive the quantized electromagnetic field in the RS basis and, within the slowly-varying amplitude approximation (SVAA), we provide a nonlinear interaction Hamiltonian for THG in ENZ nanospheres. Then, as an example of application of our formalism, we derive general expressions for the THG field, conversion efficiency, and field fluctuations. Moreover, we also consider the effect of the local field enhancement mediated by RSs on the THG generation process as a function of the input quantum state of light, including coherent, Fock, and squeezed states. Our work provides a unified analytical framework for quantum nonlinear frequency conversion in ENZ and, more generally, open-system photonic nanostructures.
%
%

Our work is organized as follows: in Sects. \ref{section2} and \ref{section3} we briefly review the quantization in lossy media, introduce the SVAA approximation, and the RSs formalism for nanocavities, respectively. Section \ref{section4} is then dedicated to the resonant basis expansion and the introduction of the RS mode operators. The general calculation of the THG field from an ENZ spherical nanocavity is then presented in Sect. \ref{section5}, while Sect.\ref{section6} discusses instead how to represent the input quantum state of the pump field in terms of RS mode operators. Then, in Sect. \ref{section7}  we derive the general expression of the THG efficiency for a spherical ENZ nanocavity, we investigate how different pump states influence its expression, and consider the simple case of a pump only coupling to dipolar RSs as a working example. Finally, conclusions and outlook are drawn in Sect. \ref{section8}.

\section{Field Quantization in Lossy Media}\label{section2}
To start with, let us briefly recall the most important results from macroscopic QED in linear, dispersive and lossy media. In the absence of external sources but presence of losses, Maxwell's equations acquire a noise-source term, proportional to the electromagnetic energy absorbed by the material as a result of light-matter interaction \cite{StefanScheel_StefanBuhmann_2008}. As a result of this, the electric field satisfies the following, inhomogeneous Helmholtz equation 
%
\begin{equation}\label{eq1}
\left[\nabla\times\nabla\times-k^{2}(\mathbf r,\omega)\right]\mathbf E(\mathbf r,\omega)=
\omega^2\mu_0\mathbf P^{(\mathrm N)}(\mathbf r,\omega),
\end{equation}
where 
\beq
k^{2}(\mathbf r,\omega)=\frac{\omega^{2}}{c^{2}}[\mathcal{\varepsilon}_1(\mathbf r,\omega) +i\mathcal{\varepsilon}_2(\mathbf r,\omega)],
\eeq
 is the (complex) wave number inside the medium, and $P^{(N)}(\vett{r},\omega)$ is the linear Langevin noise polarization  introduced to describe the energy dissipation in the absorbing dielectric medium  \cite{Scheel_2006,9yyx-s3m9,PhysRevA.58.700,PhysRevA.54.1661}. Following the procedure highlighted in Ref. \cite{vogel2006quantum}, the quantized electric field operator can then be written as
\barr\label{eq3}
\hat E_\mu(\mathbf r,\omega)&=& i\sqrt{\frac{\hbar}{\pi\varepsilon_0}}\frac{\omega^2}{c^2}\int d^3r'\sqrt{\varepsilon_2(\mathbf r',\omega)}\nonumber\\
&\times&G_{\mu\nu} (\mathbf r,\mathbf r',\omega)\hat f_\nu(\mathbf r',\omega), 
\earr
where $G_{\mu\nu}(\vett{r},\vett{r}',\omega)$ is the dyadic Green's tensor, defined as
\beq
\left[\nabla\times\nabla\times-k^{2}(\mathbf r,\omega)\right]\overleftrightarrow{\vett{G}}(\mathbf r,r',\omega)=\delta(r-r')\mathbb{I},
\eeq
where $\mathbb{I}$ is the unit dyadic, and $\hat{f}_{\nu}(\vett{r},\omega)$ are a set of noise operators obeying the standard commutation relations
\beq
\left[\hat{f}_{\mu}(\vett{r},\omega),\hat{f}_{\nu}^{\dagger}(\vett{r}',\Omega)\right]=\delta_{\mu\nu}\delta(\vett{r}-\vett{r}')\delta(\omega-\Omega),
\eeq
describing the collective, microscopic excitations of the dressed light-matter system, often called polaritons. The electric field operator in time domain is then given by
\beq
\vett{E}(\vett{r},t)=\int_0^{\infty}\,d\omega\,\vett{E}(\vett{r},\omega)\,e^{-i\omega t}.
\eeq
The fluctuation dissipation theorem then establishes a direct connection between the vacuum fluctuations of the electromagnetic field and the absorption losses of the material through the imaginary part of the dyadic Green's function as follows\cite{knoll2000qed,Scheel_2006},
\begin{eqnarray}\label{eq6}
\expectation{0}{\hat E_{\mu}(\mathbf r,\omega)\hat E_{\nu}^{\dagger}(\mathbf r',\Omega)}{0}
&=&\frac{\hbar\omega^2}{\pi\varepsilon_0c^2}\operatorname{Im}G_{\mu\nu}(\vett{r},\vett{r}',\omega)\nonumber\\
&\times&\delta(\omega-\Omega).
\end{eqnarray}
Notice, how the classical properties of the system, such as geometry, losses, mode functions, etc. are all encoded in the dyadic Green's function, which effectively serves as a bridge between the microscopic realm of dressed light-matter excitations [i.e., the polaritons described by the noise operators $\hat{f}_{\mu}(\vett{r},\omega)$], and the actual dressed photons described by $\hat{E}_{\mu}(\vett{r},\omega)$. however, the losses, encoded in the imaginary part of the Green's tensor, necessarily shape the vacuum field fluctuations.
\subsection{Slowly-Varying Amplitude Approximation}
\noindent In this work, we consider narrow-band fields centered at the carrier frequency $\omega_0$ and with a spectral width $\Delta\omega\ll\omega_0$, which allows us to apply the so-called slowly-varying amplitude approximation (SVAA) to the electric field operator in Eq. \eqref{eq3}, obtaining a description of the electric field operator compatible with standard experimental conditions \cite{Scheel_2006}. To do so, first we introduce the SVAA field as 
\beq\label{eq8}
\hat{E}_{\mu}(\vett{r},t)=\hat{\mathcal{E}}_{\mu}(\vett{r},t;\omega_0)\,e^{-i\omega_0 t}+\text{H.c.},
\eeq
where the amplitude $\hat{\mathcal{E}}_{\mu}(\vett{r},t)$ varies on a timescale much longer than the optical period $T=2\pi/\omega_0$ \cite{StefanScheel_StefanBuhmann_2008}. Then the SVAA polaritonic operator as \cite{Scheel_2006}
\begin{equation}
\hat h_{\mu}(\mathbf r,t)=\frac{1}{\sqrt{\Delta\omega}}\int_{\Delta\omega}d\omega\,\hat f_{\mu}(\mathbf r,\omega)\,e^{-i(\omega-\omega_0)t},
\end{equation}
which satisfies the equal-time commutation relations 
\beq
\left[\hat h_{\mu}(\mathbf r,t),\hat h_{\nu}^{\dagger}(\mathbf r',t)\right] =\delta_{\mu\nu} \delta(\mathbf r-\mathbf r').
\eeq
Finally, we substitute the SVAA ansatz into Eq. \eqref{eq3} to obtain the following expression for the SVAA amplitude operator
\barr\label{eq11}
\hat{\mathcal{E}}_{\mu}(\vett{r},t;\omega_0)&=&i\sqrt{\frac{\hbar\,\Delta\omega\,\varepsilon_2(\vett{r},\omega_0)}{\pi\varepsilon_0}}\frac{\omega_0^2}{c^2}\nonumber\\
&\times&\int\,d^3r'\,G_{\mu\nu}(\vett{r},\vett{r}',\omega_0)\,\hat{h}_{\nu}(\vett{r}',t).
\earr
Hereafter, for the sake of simplicity, we will omit the term ``$\omega_0$" from the argument of the SVAA electric field operator, assuming that it is implicitly understood that such field has $\omega_0$ as its carrier frequency.
\section{RESONANT STATE FORMALISM FOR A NANOCAVITY}\label{section3}
The results obtained in Eqs. \eqref{eq3} and \eqref{eq6} above are valid in general, provided that all the information about the system is encoded into the Green's tensor. A rather convenient way of representing $\overleftrightarrow{\vett{G}}(\vett{r},\vett{r}',\omega)$ for a closed, lossless resonator is to write it in terms of the normal modes of the resonator \cite{byronFuller}. For the case of lossy resonators, on the other hand, this is not possible, since losses prevent the formal definition of an orthogonal and complete set of normal modes. Instead, one uses either quasi-normal modes (QNMs) \cite{Kristensen:20}, or resonant states (RSs) \cite{Muljarov_2010}, whose complex eigen-frequencies naturally account for the finite lifetime of the cavity resonances \cite{kristensen2014modes,PhysRevA.92.053810}. Here, we focus on the latter, as they have a closed form analytical expression for the case of nanospheres. 

In general, RSs are defined as the eigensolutions of the following homogeneous Helmholtz equation

\begin{equation}
\left[\nabla\times\nabla\times  -\frac{\tilde{\omega}_n^2}{c^2}\varepsilon(\mathbf r,\tilde{\omega}_n)\right]\mathbf F_n(\mathbf r)=0,
\end{equation}
where $\tilde{\omega}_n=\omega_n-i\gamma_n$ represents the complex frequency of the RSs, $\omega_n$ being the actual resonance frequency of the mode and $\gamma_n$ the total decay rate originating from radiative leakage or material losses \cite{PhysRevA.90.013834,PhysRevA.98.033820,9yyx-s3m9}. RSs constitute a complete, orthogonal basis set, with closure relation
\beq\label{eq13}
\sum_n\varepsilon(\tilde{\omega}_n) F_{\mu,n}(\mathbf r)F_{\nu,n}(\mathbf r')=\delta_{\mu\nu}\delta(\mathbf r-\mathbf r').
\eeq
However, contrary to normal modes of Hermitian systems, RSs are normalized according to \cite{PhysRevA.90.013834,9yyx-s3m9,PhysRevA.98.033820,PhysRevB.101.045304}
\barr\label{eq9}
&2&\frac{d}{d(\omega^2)}\left[\omega^2\varepsilon(\vett{r},\omega)\right]\Big|_{\omega=\tilde{\omega}_n}\int_V dV\,\mathbf F_n(\mathbf r)\cdot\mathbf F_n(\mathbf r)\nonumber\\
&+&\frac{c^2}{\tilde{\omega}_n^2}\int_{S_V} dS\Bigg[\mathbf F_n(r)\cdot\partial_s(\mathbf r\cdot\nabla)\mathbf F_n(r)\nonumber\\
&-&(\mathbf r\cdot\nabla)\mathbf F_n(\mathbf r)\cdot \mathbf \partial_s\mathbf F_n(\mathbf r)\Bigg]=1.
\earr
In the expression above, the first integral is extended over the cavity volume $V$, while the second integral is taken over the nanocavity surface $S_V$, with $s$ representing the normal to the surface \cite{PhysRevA.90.013834,PhysRevB.101.045304}. This normalization condition includes both material dispersion and radiation leakage and provides a consistent description of the electromagnetic field in open resonant systems. Notice, moreover, that due to the fact that RSs have complex eigenfrequencies, both the normalisation condition and the closure relation above involve the square of the field, rather than its modulus square.

For the case of a nanosphere of radius $R$, the explicit, analytic expression of the RSs for TM polarized electric field is given by \cite{TMRS}
\barr\label{eq10}
\mathbf{F}_n(\mathbf r)
&=&\frac{A^{\mathrm{TM}}_{\ell}(\tilde{\omega}_n)}{k(\tilde{\omega}_n)\,r}\Bigg\{\ell(\ell+1)R_\ell Y_\ell^m(\theta,\phi)\,\hat{\mathbf r}\nonumber\\
&+&\frac{\partial}{\partial r}\left(rR_\ell\right)\nabla_\Omega Y_\ell^m(\theta,\phi)\Bigg\},
\earr
where $A_{\ell}^{TM}(\tilde\omega_n)$ is a normalisation constant, derived from Eq. \eqref{eq9}, $R_{\ell}$ is the radial mode of the RS, $Y_{\ell}^{m}(\theta,\varphi)$ are the real-valued spherical harmonics, and $\nabla_{\Omega}$ is the gradient operator in polar coordinates. The explicit expressions of these quantities are given in Appendix \ref{appendixA}. Finally, $k(\tilde\omega_n)=(\tilde\omega_n/c)\sqrt{\varepsilon(\tilde\omega_n)}$. A similar expression can be then also obtained for TE-polarized fields \cite{PhysRevA.98.033820}.

For small enough spheres, i.e., when $k(\tilde\omega_n)R\ll 1$, the eigenvalue equation determining $\tilde\omega_n$, obtained, as usual, by enforcing boundary conditions at the surface of the sphere, reduces to the familiar electrostatic regime \cite{Bohren1998}
\beq
\varepsilon(\tilde\omega_n)=-\frac{\ell+1}{\ell}.
\eeq

Using these results, we can then use the Mittag-Leffler theorem to express the Green's tensor in terms of RSs as
\cite{PhysRevB.101.045304} 
\begin{equation}\label{eq17}
G_{\mu\nu}(\mathbf r,\mathbf r',\omega)=\sum_n\frac{c^2}{\tilde{\omega}_n(\omega-\tilde{\omega}_n)}F_{n,\mu}(\mathbf r)F_{n,\nu}(\mathbf r').
\end{equation} 
\section{Resonant State expansion and operators}\label{section4}
\noindent We now have all the ingredients needed to represent the SVAA electric field operator defined in Eq. \eqref{eq11} into the basis provided by the RSs of the nanosphere. This can be in two steps: first, we can introduce the RSs into Eq. \eqref{eq11} using the expression of the Green's tensor given by Eq. \eqref{eq17}. By doing so we get
\barr\label{eq18}
\hat{\mathcal{E}}_{\mu}(\vett{r},t)&=&i\,d(\vett{r},\omega_0)\sum_n\,D_n(\omega_0)F_{n,\mu}(\vett{r})\,\hat{e}_n,
\earr
where $d(\vett{r},\omega_0)=(\omega_0^2/c^2)\sqrt{\hbar\,\Delta\omega\,\varepsilon_2(\vett{r},\omega_0)/(\pi\varepsilon_0)}$, $D_n(\omega)=c^2\sqrt{S_n}/[\tilde\omega_n(\omega-\tilde\omega_n)]$ and
\bseq\label{eq19}
\begin{align}
    \hat{e}_n(t)&\equiv\frac{1}{\sqrt{S_n}}\int\,d^3r\,F_{n,\mu}(\vett{r})\,\hat{h}_{\mu}(\vett{r},t),\\
     \hat{e}_n^{\dagger}(t)&\equiv\frac{1}{\sqrt{S_n}}\int\,d^3r\,F_{n,\mu}(\vett{r})\,\hat{h}^{\dagger}_{\mu}(\vett{r},t),
\end{align}
\eseq
are the RS mode operators, projecting the SVAA polaritonic operator $\hat{h}_{\mu}(\vett{r},t)$ onto the RS $\vett{F}_n(\vett{r})$. Notice, however, that contrary to the usual Hermitian case, here both $\hat{e}_n$ and $\hat{e}_n^{\dagger}$ are defined in terms of $\vett{F}_n(\vett{r})$ and not in terms of the mode and its conjugate. This is a consequence of the complex nature of RS, and it is required for the RS mode operators to satisfy the usual equal-time bosonic commutation relations, i.e., 
\beq
\left[\hat{e}_n,\hat{e}_m^{\dagger}\right]=\delta_{nm}.
\eeq
The explicit expression for $S_n=\int\,d^3r\,\vett{F}_n(\vett{r})\cdot\vett{F}_n(\vett{r})$ is given in Appendix A. The RS representation given above by Eq. \eqref{eq18} is the analogue of the usual normal mode expansion for Hermitian systems, but applied to the case of a lossy system. Notice, moreover, that although we have used RSs to obtain the mode expansion in Eq. \eqref{eq18}, a similar result could also be obtained using QNMs, by simply adapting the expansion coefficient $D_n(\omega)$ to match the definition of the Green's tensor for that case \cite{Kristensen:20,PhysRevLett.122.213901}.
\section{QUANTUM DESCRIPTION OF THIRD HARMONIC GENERATION}\label{section5}
\begin{figure}
    \centering
    \includegraphics[width=\linewidth]{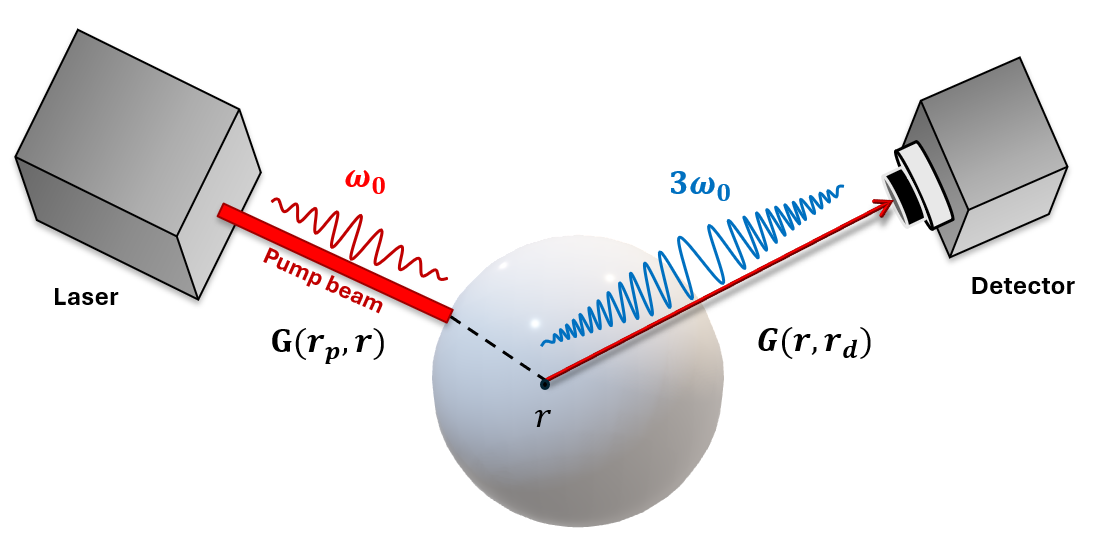}
    \caption{Schematic representation of the THG process in a spherical ENZ nanocavity. The pump field at frequency $\omega_0$ propagates from the laser source, located at position $\vett{r}_p$, to the interaction point $\vett{r}$ inside the nanocavity, described by the Green's tensor $\vett{G}(\vett{r}_p,\vett{r})$. At point $\vett{r}$, three pump photons combine to generate a new photon at frequency $3\omega_0$. The generated THG photon propagates to the detector, placed at position $\vett{r}_d$ from the nanosphere, described by the Green's tensor $\vett{G}(\vett{r},\vett{r}_d)$.}
    \label{figure1}
\end{figure}

To calculate the THG Hamiltonian and the consequent field operator, we first apply the standard nonlinear optics procedure (enabled, in this case, by the SVAA approximation) of considering the electric field as the sum of four non-overlapping modes, each centered at frequency $\omega_n$, so that $\Omega\equiv\omega_4=\omega_1+\omega_2+\omega_3$ is the energy conservation constraint imposed by the $\chi^{(3)}$-process at hand, i.e., THG. Then, using the expression of the third-order, nonlinear, causal polarization operator \cite{wilhelmi}
\begin{widetext}
    \beq\label{eq21}
\hat{P}_{\mu}^{(NL)}(\vett{r},t)=\int_{-\infty}^t\,d\tau_1\,d\tau_2\,d\tau_3\,\chi^{(3)}_{\mu\nu\sigma\lambda}(\vett{r},t-\tau_1,t-\tau_2,t-\tau_3)\,\hat{E}_{\nu}(\vett{r},\tau_1)\,\hat{E}_{\sigma}(\vett{r},\tau_2)\,\hat{E}_{\lambda}(\vett{r},\tau_3),
    \eeq
\end{widetext}
together with Eqs. \eqref{eq8} and \eqref{eq11}, and using the SVAA to take the field amplitudes $\mathcal{E}(\vett{r},\tau_k)$ out of the integral, evaluated at $\tau_k=t$, we arrive at the general expression for the THG SVAA nonlinear polarization operation, which reads
\barr
\hat{\mathcal{P}}_{\mu}(\vett{r},t;3\omega_0)&=&3\varepsilon_0\chi^{(3)}_{\mu\nu\sigma\lambda}(\vett{r},\omega_0)\hat{\mathcal{E}}_{\nu}(\vett{r},t;\omega_0)\nonumber\\
&\times&\hat{\mathcal{E}}_{\sigma}(\vett{r},t;\omega_0)\hat{\mathcal{E}}_{\lambda}(\vett{r},t;\omega_0),
\earr
where $\chi^{(3)}_{\mu\nu\sigma\lambda}(\vett{r},\omega_0)$ is the Fourier transform of the causal, third-order susceptibility appearing in Eq. \eqref{eq21} and
\beq
\hat{P}_{\mu}^{(NL)}(\vett{r},t)=\hat{\mathcal{P}}_{\mu}^{(NL)}(\vett{r},t;3\omega_0)\,e^{-i3\omega_0 t},
\eeq
holds. The THG electric field operator is then simply calculated from Eq. \eqref{eq1}, using the nonlinear polarization above as a source term, obtaining
\barr\label{eq25}
\hat{E}_{\mu}^{THG}(\vett{r},t)&=&\frac{(3\omega_0)^2}{\varepsilon_0c^2}\int\,d^3r'\,G_{\mu\nu}(\vett{r},\vett{r}',3\omega_0)\nonumber\\
&\times&\hat{P}_{\nu}^{(NL)}(\vett{r}',t).
\earr
From here, using the SVAA expression for the nonlinear polarization and the RS representation of the Green's tensor for a spherical nanocavity we obtain, after some straightforward algebra, the following expression for the THG field operator expressed in terms of RS mode operators
\barr\label{eq24}
\hat{\mathcal{E}}_{\mu}^{THG}(\vett{r},t)&=&-i\,T(\omega_0)\,\sum_{n,m,p,q}\,\mathcal{D}_{nmpq}(3\omega_0;\omega_0)\nonumber\\
&\times&V_{nmpq}\,\bar{F}_{n,\mu}(\vett{r})\,\hat{e}_m(t)\,\hat{e}_p(t)\,\hat{e}_q(t),
\earr
where
\beq
T(\omega_0)=\frac{27\omega_0^8}{c^8}\left[\frac{\hbar\,\Delta\omega\,\varepsilon_2(\omega_0)}{\pi\varepsilon_0}\right]^{3/2},
\eeq
$\mathcal{D}_{nmpq}(3\omega_0;\omega_0)=D_n(3\omega_0)D_m(\omega_0)D_p(\omega_0)D_q(\omega_0)$, and
\barr\label{eq27}
V_{nmpq}&=&\int\,d^3r\,\chi^{(3)}_{\mu\nu\sigma\lambda}(\vett{r},\omega_0)\nonumber\\
&\times&F_{n,\mu}(\vett{r})\,F_{m,\nu}(\vett{r})\,F_{p,\sigma}(\vett{r})\,F_{q,\lambda}(\vett{r}),
\earr
is the overlap integral between the pump RS and the THG RS. It is worth noticing, that in deriving Eq. \eqref{eq24} we have implicitly assumed the geometry described schematically in Fig. \ref{figure1}, i.e., that the THG field is generated in a point inside the nanosphere, where the pump field is expanded in RSs of the form $F_{n,\mu}(\vett{r})$, and then it propagates from the generation point to the detector, placed at a distance $\vett{r}_d$ from the nanosphere. Since the THG field then couples out from the nanosphere, its electric field is primarily described by the RSs \emph{outside} the sphere, i.e., $\bar{F}_{n,\mu}(\vett{r})$. To make Eq. \eqref{eq24} consistent with Fig. \ref{figure1}, therefore, one should set $\vett{r}=\vett{r}_d$ in the argument of the field operator, i.e., $\hat{\mathcal{E}}_{\mu}^{THG}(\vett{r},t)\,\rightarrow\,\hat{\mathcal{E}}_{\mu}^{THG}(\vett{r}_d,t)$. To avoid making the notation heavier than it already here, we drop the subscript $d$, assuming that this is implicitly understood. The fully correct notation can be then retrieved by employing the substitution $\vett{r}\,\rightarrow\,\vett{r}_d$ in the argument of any barred RS $\bar{\vett{F}}(\vett{r})$.
\section{Quantum state of the pump field in RS mode basis}\label{section6}
\noindent To calculate the THG efficiency we need to specify the quantum state of the impinging pump beam. A convenient way to do this, especially since we then want to evaluate the effect of different pump states on the THG efficiency, is to represent the state of the input field in terms of RS mode operators, so that its projection onto RSs is automatically accounted for. This is a particularly convenient method when it comes to coherent and squeezed pump states, as it will make the evaluation of the expectation value of the THG efficiency much easier to calculate. 

To this aim, let us assume the pump field to be in a SVAA mode $\vett{E}_p(\vett{r},t;\omega_p)=\vett{u}_p(\vett{r},\omega_p)\,\exp(-i\omega_p t)$ and introduce the pump operator
\beq
\hat{A}_p^{\dagger}=\frac{1}{\sqrt{V}}\sum_n\,C_n(\omega_p)\,\hat{e}^{\dagger}_n,
\eeq
(and a similar expression for the pump annihilation operator) where the coefficients $C_n(\omega_p)=\int\,d^3r\,\vett{F}_n(\vett{r})\cdot\vett{u}_p(\vett{r},t)$ is the coupling coefficient of the pump field with each RS of the nanosphere. Then, the input pump state for the case of a Fock, coherent, and squeezed state can be simply written as 
$\ket{n_p}=(\hat{A}_p^{\dagger})^n/\sqrt{n!}\ket{0}$, $\ket{\alpha_p}=\hat{\mathcal{D}}(\alpha)\ket{0}$, and $\ket{\xi_p}=\hat{S}(\xi)\ket{0}$, respectively, where $\hat{\mathcal{D}}(\alpha)$ and $\hat{S}(\xi)$ are the usual displacement and squeezing operators, defined with respect to the pump operators $\hat{A}_p$ and $\hat{A}_p^{\dagger}$ \cite{loudon2000quantum,liscidini}.

The main advantage of introducing the pump operator in this way comes from the fact that when evaluating expectation value of field operators with respect to either coherent or squeezed states, the following identities can be used to simplify the calculations
\bseq
\begin{align}
    \hat{\mathcal{D}}^{\dagger}(\alpha)\,\hat{e}_n\,\hat{\mathcal{D}}(\alpha)&=\hat{e}_n+\frac{\alpha}{\sqrt{V}}\,C_n(\omega_0),\\
    \hat{S}^{\dagger}(\xi)\,\hat{e}_n\,\hat{S}(\xi)&=\hat{e}_n+\frac{1}{\sqrt{V}}C_n(\omega_0)\Bigg[(\operatorname{cosh}r-1)\,\hat{A}_p\nonumber\\
    &-e^{i\theta}\operatorname{sinh}r\,\hat{A}_p^{\dagger}\Bigg],
\end{align}
\eseq
where $\xi=r\,\exp(i\theta)$.

Using these expressions, we can for example define the classical pump field in terms of the RSs as the expectation value of the SVAA field operator in Eq. \eqref{eq18}  with respect to the coherent pump state $\ket{\alpha_p}$, obtaining
\barr\label{eq30}
\expectation{\alpha_p}{\hat{\mathcal{E}}_{\mu}(\vett{r},t)}{\alpha_p}&=&\expectation{0}{\hat{\mathcal{D}}^{\dagger}(\alpha)\,\hat{\mathcal{E}}_{\mu}(\vett{r},t)\,\hat{\mathcal{D}}(\alpha)}{0}\nonumber\\
&=&i\,\alpha\,\frac{\omega_p^2}{c^2}\,\sqrt{\frac{\hbar\Delta\omega\varepsilon_2(\vett{r},\omega_p)}{\pi\varepsilon_0V}}\sum_nD_n(\omega_p)\nonumber\\
&\times&C_n(\omega_p)F_{n,\mu}(\vett{r})\nonumber\\
&\equiv&\alpha\,E_0\,u_{p,\mu}(\vett{r},t),
\earr
where $\vett{u}_p(\vett{r},t)$ is the classical mode field of the pump, and $E_0=\sqrt{\hbar\Delta\omega/(\pi\varepsilon_0V)}$ is the single photon amplitude.
\section{Third-Harmonic Generation Efficiency}\label{section7}
We define the THG efficiency as the ratio between the intensity of the THG signal and the intensity of the impinging pump as
\beq\label{eq32I}
\eta_{THG}=\frac{I_{THG}(\vett{r};3\omega_p)}{I_p(\vett{r};\omega_p)},
\eeq
where the intensity of both THG and pump fields is calculated as the expectation value of the number operator over the state of the input field, i.e., \cite{loudon2000quantum}
\beq
I(\vett{r};\omega)=2\varepsilon_0\,c\,n(\omega)\expectation{\psi_{in}}{\hat{\mathcal{E}}_{\mu}^{\dagger}(\vett{r},t)\hat{\mathcal{E}}_{\mu}(\vett{r},t)}{\psi_{in}},
\eeq
with $n(\omega)=\operatorname{Re}\sqrt{\varepsilon(\omega)}$ the refractive index of the nanosphere. The general expression of the THG intensity can be written in the form
\beq\label{eq33I}
I_{THG}(\vett{r};3\omega_p)=G_{THG}(\omega_p\,)\mathcal{I}(\vett{r}; \omega_p),
\eeq
where
\beq\label{eq40}
G_{THG}(\omega_p)=2\varepsilon_0\,c\,n(3\omega_p)T(\omega_p)^2N_{THG},
\eeq
and
\barr\label{eq35}
\mathcal{I}(\vett{r};\omega_p)&=&\Bigg|\sum_{n,m,p,q}\mathcal{D}_{nmpq}(3\omega_p;\omega_p)V_{nmpq}\nonumber\\
&\times&C_m(\omega_p)C_p(\omega_p)C_q(\omega_p)\,\bar{F}_{n,\mu}(\vett{r})\Bigg|^2,
\earr
where $N_{THG}$ is given by
\beq
N_{THG}=\begin{cases}
|\alpha|^6, & \ket{\psi_{in}}=\ket{\alpha_p},\\
n(n-1)(n-2), & \ket{\psi_{in}}=\ket{n_p},\\
\operatorname{sinh}^4r\,\operatorname{cosh}^2r, & \ket{\psi_{in}}=\ket{\xi_p}.
\end{cases}
\eeq
If we then call $I_p=N_p\,I_0(\vett{r};\omega_p)$ the intensity of the pump field, where $N_p$ given by
\beq
N_p=\begin{cases}
|\alpha|^2, & \ket{\psi_{in}}=\ket{\alpha_p},\\
n, & \ket{\psi_{in}}=\ket{n_p},\\
\operatorname{sinh}^2r, & \ket{\psi_{in}}=\ket{\xi_p},
\end{cases}
\eeq
and $I_0(\vett{r};\omega_p)=2\varepsilon_0cn(\omega_p)|\vett{u}_p(\vett{r},t)|^2$, we can then write the THG efficiency in the following, compact form
\barr\label{eq38}
\eta_{THG}=\left[\frac{n(3\omega_p)}{n(\omega_p)|\vett{u}_p|^2}\right]\frac{T(\omega_p)^2}{V^2}\frac{N_{THG}}{N_p}\mathcal{I}(\vett{r};\omega_p).
\earr
This is the first result of our work. By combining the Green's function quantization framework and the RSs expansion, we have derived the expression of the quantized electric field inside an ENZ nanosphere using RS mode operators $\hat{e}_n$. Consequently, we have calculated, for the case of THG, the nonlinear electric field operator and the analytic expression for the THG efficiency, for the general case of an arbitrarily shaped impinging pump field, covering the case of the pump field being initially in a Fock, coherent, or squeezed state.

\subsection{A simple example}
We now consider a simple example, namely an ENZ nanosphere of radius $R=20$ nm, assumed to host only dipolar RSs (enabled by ensuring that $k(\tilde\omega_n)R\ll1$ holds), derive the far field expression of the generated THG field, and calculate the THG efficiency for different input quantum states of the pump. As ENZ material we consider ITO, and use the Drude model
\beq
\varepsilon(\omega)=\varepsilon_b-\frac{\omega_p^2}{\omega(\omega+i\gamma)},
\eeq
to describe its permittivity, where $\varepsilon_b=3.8$ is the background permittivity, $\omega_p=3\times 10^{15}$ Hz is the plasma frequency, and $\gamma=1.91\times 10^{14}$ Hz is the damping parameter \cite{9yyx-s3m9}. Moreover, we assume the third-order susceptibility tensor to be isotropic and equal to $\chi^{(3)}=\chi^{(3)}_{xxxx}=5.2\times 10^{-17}$ $\text{V}^2/\text{m}^2$ \cite{10.1063/5.0240990}. For the pump field, we choose a plane-wave, TM-polarized mode function $\vett{u}_p(z,t)=E_0\,\exp[i(k_0z-\omega_p t)]\uvett{x}$ propagating along the $z$-direction. From Eq. \eqref{eq10}, setting $\ell=1$ gives us the dipolar RSs of the nanosphere as 
\bseq\label{eqsF1}
\begin{align}
\vett{F}_1^{-1}(\vett{r})&=-\sqrt{\frac{2S_1}{V}}\left[\sin\varphi\left(\sin\theta\,\uvett{r}+\cos\theta\,\hat{\boldsymbol\theta}\right)+\cos\varphi\,\hat{\boldsymbol\varphi}\right],\\
\vett{F}_1^{0}(\vett{r})&=\sqrt{\frac{2S_1}{V}}\left[\cos\theta\,\uvett{r}-\sin\theta\,\hat{\boldsymbol\theta}\right],\\
\vett{F}_1^{1}(\vett{r})&=\sqrt{\frac{2S_1}{V}}\left[-\cos\varphi\left(\sin\theta\,\uvett{r}+\cos\theta\,\hat{\boldsymbol\theta}\right)+\sin\varphi\,\hat{\boldsymbol\varphi}\right],
\end{align}
\eseq
where the superscripts $\{0,\pm1\}$ refer to the possible values of the quantum number $m\in[-\ell,\ell]$, $V=4\pi/3R^3$ is the volume of the nanosphere, and the explicit expression of $S_1$ is given in Appendix A. Expanding the input pump field onto the RSs gives using $\vett{u}_p=\sum_nC_n(\omega)\vett{F}_n$ gives us the following expression for the expansion coefficients
\bseq\label{eq41}
\begin{align}
C_1^0(\omega)&=0,\\
C_1^{-1}(\omega)&=0,\\
C_1^1(\omega)&=\frac{20\,\pi\,R}{3k_0^2\,\sqrt{S_1V}}f(k_0R),
\end{align}
\eseq
with $f(x)=\sin x-x\cos x\simeq x^3/3$ for $x\ll 1$. The detailed derivation of the expansion coefficients are reported in appendix B. 

From Eq. \eqref{eq40}, we see that since the input field only excited a dipolar RS, the THG signal will also be generated in a dipolar RS, aligned with that of the pump signal. This essentially means, that the choice $(\ell=1,m=1)$ holds for both the pump and the THG field. Using this fact into Eq. \eqref{eq27} we can write the overlap integral as
\barr
V_{1111}=-\frac{4S_1^2}{V}\chi_{xxxx}^{(3)}.
\earr
Imposing $n=m=p=q=1$ in Eq. \eqref{eq35} gives, after a little algebra, the following expression for the THG efficiency
\beq\label{eq43}
\eta_{THG}^{(dipole)}(\vett{r})=\eta_0\,\eta_{cav}\,\eta_q\,\left|\bar{\vett{F}}_1^1(\vett{r})\right|^2,
\eeq
where
\beq
\eta_{cav}=\frac{|S_1|^5f^6(k_0R)\,|\chi^{(3)}|^2}{V^5\,|\tilde\omega_1^4\gamma_1^3(2\omega_p+i\gamma_1)|^2},
\eeq
is the efficiency enhancement factor due to the cavity, $\eta_q=N_{THG/N_p}$ is the quantum efficiency,
\beq
\eta_0=\frac{576\times 10^6\pi^2n(3\omega_p)\hbar^2\,\Delta\omega^2\,\omega_p^4\,c^{12}\varepsilon_2^3(\omega_p)}{\varepsilon_0^2\,n(\omega_p)},
\eeq
and $\bar{\vett{F}}_1^1(\vett{r})$ is the $(\ell=1,m=1)$ RS outside the nanosphere, and we have implicitly assumed the pump field to be in resonance with the RS, i.e., $\omega_p=\omega_1$. 

The explicit expression for the RSs $\bar{\vett{F}}_1^m(\vett{r})$ is obtained by using the expression of the radial function $R_{\ell}$ outside the cavity (see Appendix A), which, in the far field limit $r\,\rightarrow\,\infty$, and for $k(\omega_1)R\ll 1)$, reads
\bseq
\begin{align}
    \bar{\vett{F}}_1^{-1}(\vett{r})&=-i\sqrt{\frac{2S_1}{V}}\frac{k^2(\omega_1)R^3}{r}\Bigg[\nonumber\\
    &\cos\theta\,\sin\varphi\,\hat{\boldsymbol\theta}+\cos\varphi\,\hat{\boldsymbol\varphi}\Bigg],\\
    \bar{\vett{F}}_1^0(\vett{r})&=-i\sqrt{\frac{2S_1}{V}}\frac{k^2(\omega_1)R^3}{r}\sin\theta\,\hat{\boldsymbol\theta},\\
    \bar{\vett{F}}_1^1(\vett{r})&=-i\sqrt{\frac{2S_1}{V}}\frac{k^2(\omega_1)R^3}{r}\Bigg[\nonumber\\
    &\cos\theta\,\cos\varphi\,\hat{\boldsymbol\theta}-\sin\varphi\,\hat{\boldsymbol\varphi}\Bigg].
\end{align}
\eseq
For a pump pulse resonant with the ENZ frequency, i.e., $\omega_p=\omega_{ENZ}=1.527\times 10^{15}$ Hz, corresponding to $\lambda_{ENZ}=1234$ nm, and with duration $\Delta\tau=150$ fs, one obtains, in the undepleted pump approximation and assuming $|\alpha|^2\approx 10^9$ an overall total efficiency $\eta_{THG}=\int\,d^3r\,\eta_{THG}^{(dipole)}(\vett{r})$ integrated over all space (and normalized to the nanosphere-detector distance) of about
\beq
\eta_{THG}\approx 5.5\times 10^{-8},
\eeq
which is about two orders of magnitude higher than the one measured for ITO slabs \cite{10.1063/5.0240990}. Notice, moreover, that for this particular example, the enhanced single photon electric field amplitude is about $E_0=\sqrt{\hbar\Delta\omega/(\pi\varepsilon_0 V)}\approx 2\times 10^6$.

Realistically, however, the detector will subtend a solid angle $\Omega_D$ as seen by the nanosphere, and only a fraction of the THG intensity scattered by the nanosphere will enter the detector and be measured. A more realistic value for the THG efficiency would then be
\beq
\eta_{THG}^{(av)}=\int_{\Omega_D}\,d\Omega\,\eta_{THG}^{(dipole)}(\vett{r}).
\eeq
For a circular detector of radius $\rho$ placed at a distance $D\gg\rho$ from the nanosphere, the subtended solid angle is $\omega_D=2\pi(1-\cos\theta_D)\simeq\pi\theta_D^2$, where $2\theta_D=\arctan(\rho/D)$ is the tip angle of the cone subtended by the detector. In this limit, the integral above simply becomes
\beq
\eta_{THG}^{(av)}\approx\frac{\Omega_D}{2}\eta_{THG}\approx 8 \times 10^{-10},
\eeq
assuming an acceptance solid angle of the detector $\Omega_D/4\pi=2.5\times 10^{-3}$, which is in line with earlier measurements \cite{10.1063/5.0240990}.

\subsection{Quantum fluctuations of the THG field}
\noindent In this section, we focus on the study of the statistical properties of the THG radiation generated by the ENZ nanosphere, by looking at the field fluctuations of the THG field and how they link to fluctuations of the pump field. These quantities, in fact, provide insight on the quantum nature of the nonlinear conversion process. The variance of the third-harmonic electric field is defined as \cite{gerry2005introductory}
\begin{equation}
(\Delta E_{\mu}^{THG})^2=\left\langle[\hat{E}_\mu^{THG}(\mathbf r,t)]^2\right\rangle-
\left|\left\langle \hat E_\mu^{THG}(\mathbf r,t)\right\rangle\right|^2,
\end{equation}
with the second term evaluating to zero for Fock and squeezed (vacuum) states. After some simple, but cumbersome, algebra, one can write the general expression of the fluctuations of the field in Eq. \eqref{eq25} as
\beq
\left(\Delta\,\hat{\mathcal{E}}_{\mu}^{THG}(\vett{r},t)\right)^2=|\Gamma|^2\sigma(n)+\beta(n,\Gamma),
\eeq
where
\barr\label{eq49}
\Gamma&=&\frac{T(\omega_p)}{V^{3/2}}\sum_{n,m,p,q}\mathcal{D}_{nmpq}(3\omega_p;\omega_p)V_{mnpq}\bar{F}_{n,\mu}(\vett{r})\nonumber\\
&\times&C_m(\omega_p)C_p(\omega_p)C_q(\omega_p),
\earr
is the cavity enhancement factor, 
\beq
    \sigma(n)=\begin{cases}
        3(3|\alpha|^4+6|\alpha|^2+2), & \ket{\psi_{in}}=\ket{\alpha_p},\\
        2n^3+3n^2+13n+6, & \ket{\psi_{in}}=\ket{n_p},\\
        3\operatorname{sinh}^2r\left(\operatorname{sinh}^2+\frac{3}{2}\right)^2+6, & \ket{\psi_{in}}=\ket{\xi_p},
    \end{cases}
\eeq
and
\beq
\beta(n,\Gamma)=\begin{cases}
    0, & \ket{\psi_{in}}=\ket{\alpha_p},\\
    0, & \ket{\psi_{in}}=\ket{n_p},\\
    \frac{15}{4}\operatorname{sinh}^32r\,\operatorname{Re}\left[\Gamma^2e^{i3\theta}\right], & \ket{\psi_{in}}=\ket{\xi_p},
\end{cases}
\eeq
are the variance parameters for the various input quantum states. Similarly, the variance for the impinging field can be expanded into the RSs of the nanosphere, and written in a similar expression to EQ. \eqref{eq49} as
\beq\label{eq55}
\left(\Delta\,\hat{\mathcal{E}}_{\mu}^{pump}(\vett{r},t)\right)^2=|\Lambda_{\mu}|^2\sigma_p(n)+\beta_p(n,\Lambda),
\eeq
where now
\beq
\Lambda_{\mu}=\frac{d(\vett{r},\omega_p)}{\sqrt{V}}\sum_nD_n(\omega_p)F_{n,\mu}(\vett{r})C_n(\omega_p),
\eeq
and with the variance parameters
\beq
\sigma_p(n)=\begin{cases}
    1, & \ket{\psi_{in}}=\ket{\alpha_p},\\
    2n+1, &  \ket{\psi_{in}}=\ket{n_p},\\
    2n+1 & \ket{\psi_{in}}=\ket{\xi_p},
\end{cases}
\eeq
and $\beta_p(n,\Lambda)=\operatorname{sinh}r\,\operatorname{Re}\left[\Lambda^2\exp(i\theta)\right]$, which is nonzero only for squeezed states.

From here, one can then evaluate the relative field fluctuations, defined as the ratio between the fluctuations of the THG field and the fluctuations inherently possessed by the pump field as
\beq
\delta=\frac{\Delta\,\hat{\mathcal{E}}_{\mu}^{THG}}{\Delta\,\hat{\mathcal{E}}_{\mu}^{pump}}=\sqrt{\frac{|\Gamma_{\mu}|^2\sigma(n)+\beta(n)}{|\Lambda_{\mu}|^2\sigma_p(n)+\beta_p(n)}}.
\eeq
This is the second result of our work. Equations \eqref{eq49} and \eqref{eq55} give the general expression of the variance of the THG and pump field, respectively, for arbitrary input pump state (and spatial mode), expanded onto the RSs of the ENZ nanosphere. The coefficients $C_n(\omega_p)$ appearing in the expressions of $\Gamma_{\mu}$ and $\Lambda_{\mu}$ account for the overlap between the spatial mode of the pump and the RSs. Moreover, the selection rules regulating the RSs in which the THG field is generated (regulated by the nonlinear process, i.e., by the third-order susceptibility tensor $\chi_{\mu\nu\sigma\lambda}^{(3)}$) are automatically taken into account by the expansion through the overlap integral $V_{nmpq}$.
\section{CONCLUSION}\label{section8}
In this work, we have applied the Green's tensor quantization formalism typical of macroscopic QED to the case of an ENZ nanosphere, described within the framework of RSs. Taking THG as a working example, we have derived general expressions for the THG generation efficiency and THG field fluctuations, expressed in terms of the RSs of the nanosphere. To benchmark our results, we have calculated the THG generation efficiency for the simple case of an ENZ nanosphere supporting only dipolar RSs, obtaining a value for $\eta_{THG}$ in accordance with experimental measurements made on ITO described by the same model parameters as those we used. 

Although here we have worked with RSs and nanospheres, the main results of our work, i.e., Eqs.  \eqref{eq25}, \eqref{eq38}, \eqref{eq49}, and \eqref{eq55} are valid for any geometry and any modal approach to nanocavities, provided that the quantities appearing in them (e.g., the functions $\vett{F}_n$ and $\bar{\vett{F}}_n$) are interpreted accordingly. In particular, adopting a QNMs approach would allow the study of more complicated geometries than a nanosphere, that are in general not accessible analytically. Similarly, our framework can also be extended to other nonlinear processes than THG. This can be done by replacing Eq. \eqref{eq21} with a definition of the causal nonlinear polarizaiton describing the nonlinear process of interest, and then use this new definition in Eq. \eqref{eq25} to generate the correspondent nonlinear electric field operator.

The framework developed in this work constitutes an important tool for both theoretical and experimental investigation of nonlinear effects in ENZ nanostructures, as it gives explicit expressions for the measurable quantities, such as intensities and field fluctuations, in terms of the RSs (or, more generally, QNMs) of the nanocavity, and the electromagnetic field interacting with them. 
\section*{Acknowledgements}
The authors acknowledge the financial support from the Research Council of Finland Flagship Program (PREIN - decision Grant No. 320165). S.A. also acknowledges the support from the I-DEEP doctoral pilot program.

\section*{Appendix A: Explicit expression of the Resonant States for a nanosphere}\label{appendixA}
%
In this Appendix, we provide explicit expresisons for the radial ($R_{\ell}$) and angular ($Y_{\ell}^m(\theta, \varphi)$) functions composing the RS of a nanosphere, as well as the explicit expression for the normalisation constant $A_{\ell}^{TM}(\tilde\omega_n)$. 

Inside the nanocavity, the radial function $R_{\ell}$ can be express in terms of spherical Bessel functions of the first kind $j_{\ell}(x)$, while the boundary conditions for outgoing waves at infinity imposes to choose spherical Hankel functions of the first kind $h_{\ell}^{(1)}(x)$ outside the sphere \cite{Olver2010}. Taking this into account $R_{\ell}$ can be written as follows
\begin{equation}
R_{\ell}=
\begin{cases}
\dfrac{j_{\ell}\!\left(k({\tilde{\omega}_n)}\,r\right)}{j_{\ell}\!\left(k{(\tilde{\omega}_n)}R\right)},
& r<R,
\\[12pt] \dfrac{h_{\ell}^{(1)}\!\left(k(\omega_n)r\right)}{h_{\ell}^{(1)}\!\left(k(\omega_n)R\right)},
&r>R,
\end{cases}
\label{eq:radial_function}
\end{equation}
where $k(\tilde\omega_n)=(\tilde\omega_n/c)\sqrt{\varepsilon(\tilde\omega_n)}$ and $k(\omega_n)=\omega_n/c$, respectively, represent the wavenumbers inside and outside the nanocavity.
 
The angular component of the cavity RSs are instead described by real-valued spherical harmonic $Y_{\ell}^m(\theta,\phi)$, defined as \cite{PhysRevA.90.013834}
\beq
Y_{\ell}^m(\theta,\varphi)=\sqrt{\frac{(2\ell+1)(\ell-m)!}{2\pi(\ell+m)!}}P_{\ell}^m(\cos\theta)\zeta_m(\varphi),
\eeq
where $P_{\ell}^m(\cos\theta)$ are the associated Legendre polynomials \cite{byronFuller}, and
\beq
\zeta_m(\varphi)=\begin{cases}
\sin m\varphi, & m<0,\\
\frac{1}{\sqrt{2}}, & m=0,\\
\cos m\varphi, & m>0.
\end{cases}
\eeq
Moreover, the following identities hold
\bseq
\begin{align}
    \nabla_{\Omega}Y_{\ell}^m(\theta,\phi)&=\left[\frac{\partial}{\partial\theta}\hat{\boldsymbol\theta}+\frac{1}{\sin\theta}\frac{\partial}{\partial\phi}\hat{\boldsymbol\phi}\right]Y_{\ell}^m(\theta,\phi),\\
    \nabla_{\Omega}^2Y_{\ell}^m(\theta,\phi)&=-\ell(\ell+1)Y_{\ell}^m(\theta,\varphi).
\end{align}
\eseq

FInally, according to the normalization condition given in Eq. \eqref{eq9}, the explicit expression of the normalization coefficient $A_{\ell}^{\mathrm{TM}}(\tilde{\omega}_n)$ appearing in Eq. \eqref{eq10} can be written as
\begin{equation}
    A_l^{\mathrm{TM}}(\tilde{\omega}_n)=\left[\ell (\ell+1)R^{3}\Big(\varepsilon(\tilde{\omega}_n)-1\Big)\varepsilon(\tilde{\omega}_n)L_{\ell}(\mathrm{\tilde{\omega}_n})\right]^{-1/2},
\end{equation}
where \(L_{\ell}(\tilde\omega_n)\), contains the geometrical properties of the nanocavity and the dispersive response of the ENZ material via
\barr
L_{\ell}(\omega)&=&
\frac{1}{\varepsilon(\omega)} \frac{j_{\ell-1}(X)}{j_{\ell}(X)}-\frac{1}{X}\nonumber\\
&-&\frac{\ell}{X^{2}}+D_n M_{\ell}(\omega),
\earr
where $X=k(\tilde\omega_n)R$, $M_{\ell}(\omega)$ is given by
\barr
M_{\ell}(\omega)&=&\frac{1}{\varepsilon(\omega)-1}\Biggl[\frac{j_{\ell-1}^{2}(X)}{j_{\ell}^{2}(X)}-\frac{j_{\ell-2}(X)}{j_{\ell}(X)}\nonumber\\
&-&\frac{2\ell}{X^{2}}\Biggr],
\earr
and 
\begin{equation}
D_n=\frac{\omega}{2\varepsilon(\omega)}\left[\frac{\partial \varepsilon(\omega)}{\partial \omega}\right]\Bigg|_{\omega=\tilde{\omega}_n}.
\end{equation}
Moreover, in the quasi-static limit where $X\ll 1$, it can be shown \cite{9yyx-s3m9} that
\beq\label{eq80}
\int_V\,d^3r\,\vett{E}_n(\vett{r})\cdot\vett{E}_m(\vett{r})=\frac{\delta_{nm}}{\tilde\omega_n\varepsilon'(\tilde\omega_n)}\equiv\delta_{nm}S_n,
\eeq
where prime denotes derivative with respect to the argument.


\section*{Appendix B: Derivation of the Expansion Coefficients $C_n(\omega)$}\label{appendixB}
\noindent In this Appendix, we derive the general expression for the expansion coefficient $C_n(\omega)$ representing the projection of an impinging field onto the RSs of the nanosphere, in the quasi-static limit $k(\tilde\omega_n)R\ll1$. For the incident field, we consider a monochromatic, $x$-polarized emode function propagating along the $z$ direction and defined as
\begin{equation}
\mathbf u_{pump}(z,t)
=
e^{i(k_0 z-\omega_0 t)}\,\hat{\mathbf x},
\end{equation}
and represent it in the RS basis as \cite{Muljarov_2010,PhysRevA.90.013834} 
\begin{equation}
\mathbf u_{pump}(z,t)=e^{-i\omega_0 t}\sum_n C_n(\omega)\,\mathbf F_n(\mathbf r),
\end{equation}
where $k_0=\omega/c$  is the wave vector of the incident field in vacuum. From here, we can extract the explicit expression of the expansion coefficient using the orthogonality relation for RS, i.e., Eq. \eqref{eq80}, to get 
\begin{equation}
C_n(\omega)=\frac{1}{S_n}\int_V d^3r\,\mathbf F_n(\mathbf r)\cdot\mathbf u_{pump}(z,t).
\end{equation}
Using the Rayleigh identity in spherical coordinates, with $z=r\cos\theta$ \cite{byronFuller}
\begin{equation}\label{eq:RayleighAppendix}
e^{ik_0r\cos\theta}=\sum_{l=0}^{\infty}i^l(2l+1)j_l(k_0r)P_l(\cos\theta),
\end{equation}
and using the explicit expresison of the RSs given by Eq. \eqref{eq10}, we can write the general expression of the expansion coefficients as follows
\barr\label{eq85}
C_n(\omega)&=&\frac{\pi A_n^{TM}(\tilde\omega_n)g_{\ell}}{S_n}\sum_s\,i^s(2s+1)\nonumber\\
&\times&\left[I_{\ell,s}^{(1)}I_r^{(1)}+I_r^{(2)}\Big(I_{\ell,s}^{(2)}+I_{\ell,s}^{(3)}\Big)\right],
\earr
where $g_{\ell}=\sqrt{(2\ell+1)(\ell-1)!/[2\pi(\ell+1)!]}$, and the quantities $I_{r,\theta}^{(n)}$ are given as follows
\bseq
\begin{align}
    I_{\ell,s}^{(1)}&=-\frac{4\ell\,\delta_{s,\ell-1}}{(2\ell-1)(2\ell+1)}+\frac{4(\ell+1)\,\delta_{s,\ell+1}}{(2\ell+1)(2\ell+3)},\\
    I_{\ell,s}^{(2)}&=-\frac{\ell(\ell+1)\,\delta_{s,\ell-1}}{(2\ell-1)(2\ell+1)}+\frac{\ell(\ell+1)\,\delta_{s,\ell+1}}{(2\ell+1)(2\ell+3)},\\
    I_{\ell,s}^{(3)}&=\begin{cases}
        -2, & s<\ell, \text{and}\,s+\ell\,\text{odd},\\
        0, & \text{otherwise},
    \end{cases}
\end{align}
\eseq
and 
\bseq
\begin{align}
    I_r^{(1)}&=\frac{1}{k(\tilde\omega_n)}\int_0^R\,dr\,r\,j_s(k_0r)\,R_{\ell}(r),\\
    I_r^{(2)}&=\frac{1}{k(\tilde\omega_n)}\int_0^R\,dr\,r\,j_s(k_0r)\frac{d}{dr}\left[r\,R_{\ell}(r)\right],
\end{align}
\eseq
for the radial integrals. In general, these integrals do not have closed form solutions, and can only be evaluated numerically. However, in the quasi-static limit, the radial part of the RSs can be approximated as
\bseq
\begin{align}
    \frac{1}{k(\tilde\omega_n)r}R_{\ell}(r)&\simeq\frac{r^{\ell-1}}{k(\tilde\omega_n)R},\\
    \frac{1}{k(\tilde\omega_n)r}\frac{d}{dr}\left[r\,R_{\ell}(r)\right]&\simeq(\ell+1)\frac{r^{\ell-1}}{k(\tilde\omega_n)R},
\end{align}
\eseq
and the integrals above admit closed form solutions in terms of regularized hypergeometric functions, which, for $\ell=1$ reduce to
\beq
    I_r^{(1)}=I_r^{(2)}/2=\frac{\sin k_0R-k_0R\cos k_0R}{k_0^3}.
\eeq
In the quasi-static limit, and for $\ell=1$, moreover, the normalization coefficient $A_1^{TM}(\tilde\omega_1)$ can be approximated as
\beq
A_1^{TM}(\tilde\omega_1)\simeq i\sqrt{\frac{k^2(\tilde\omega_1)S_1}{2R}}.
\eeq
Substituting all of this into Eq. \eqref{eq85} finally gives $C_1^0(\omega)=C_1^{-1}(\omega)=0$, and
\beq
C_1^{1}(\omega)=\frac{20\,\pi}{3\sqrt{S_1V}}\frac{R}{k_0^2}\frac{(k_0R)^3}{3},
\eeq
where we used the fact that $k(\tilde\omega_1)=k_0\sqrt{\varepsilon(\tilde\omega_1)}=ik_0\sqrt{2}$, since $\varepsilon(\tilde\omega_1)=-2$ for $\ell=1$, and $\left[\sin x-x\,\cos\,x\right]\simeq x^3/3$. Notice, that since the impinging beam is $x$-polarized, only the dipole RS aligned along the $x$-direction (corresponding to the choice $m=1$) will contribute to the expansion of the input field into RSs, as expected.

\bibliography{references}

\end{document}